# A New Noncentrosymmetric Superconducting Phase in the Li-Rh-B System

Hiroyuki Takeya[1,3] *, Hiroki Fujii[1], Mohammed Elmassalami[2], Francisco Chaves[2], Shuuichi Ooi[1], Takashi Mochiku[1], Yoshihiko Takano[1,3], Kazuto Hirata[1] and Kazumasa Togano[1,3]

[1] *National Institute for Materials Science, Tsukuba 305-0047, Japan*
[2] *Instituto de Fisica, Universidade Federal do Rio de Janeiro, Caixa Postal 68528, 21945-970 Rio de Janeiro, Brazil*
[3] *JST, Transformative Research-Project on Iron Pnictides (TRIP), Chiyoda-ku, Tokyo 102-0075, Japan*

Superconductivity, at 2-3 K, was observed in a novel phase of the ternary Li-Rh-B system. The structural phase exhibits a large noncentrosymmetric cubic unit cell with the $a$-parameter being within $1.208 \leq a \leq 1.215$ nm. This phase is stable over a wider compositional range of $Li_xRhB_y$ ($0.6 < x < 2, 1 < y < 2$). The superconductivity, as well as the unit cell volume, is sensitive to the Li/B content but it is manifested with $T_c \geq 1.8$ K over a wider compositional range: the highest $T_c \simeq 3$ K occurs for $x:y \approx 0.9:1.5$ with $a \simeq 1.209$ nm. The superconducting shielding fraction of most samples is almost 80% of that of Sn. The lower critical field, $H_{c1}(0)$, is $\sim 65$ Oe while the upper one, $H_{c2}(0)$ is determined from extrapolation to be higher than 14 kOe. We discuss the influence of pressure on $T_c$ and also the influence of the lack of inversion symmetry on the superconducting properties.

A discovery of a novel superconducting family is usually accompanied by a surge of intense research activities in the field of solid-state physics and material science; such was the case after the recent discovery of superconductivity in the iron pnictide (e.g. $LaO_{1-x}F_xFeAs$[1] and $Ba_{1-x}K_xFe_2As_2$[2]) and in the noncentrosymmetry, nonCS, structures (such as $CePt_3Si$[3] and $Li_2Pt_3B$[4–6]). While the interest in the former types is related to their unconventional character (most probably associated with magnetic fluctuation), the latter types of superconductors attract much scientific attention due to the unique and interesting superconducting properties that arise in crystal structures with no inversion centers.[7,8]

The presence of an antisymmetric spin-orbit coupling, ASOC, in such noncentrosymmetric structure would lift the spin degeneracy of the electron bands and as such Cooper-pair states can not be classified into even (spin-singlet) and odd (spin-triplet) parity. Rather, an admixture of parity states may occur and the structure of the quasiparticle gap function may contain line or point nodes.[8] In that case, most of the thermodynamical properties within such a superconducting state would reveal unconventional character such as power-law thermal dependence.[7] Such features are manifested in a number of compounds e.g. CePt$_3$Si (space group $P4mm$, $T_c$= 0.75 K),[3] UIr (space group $P2_1$, $T_c \approx$ 0.15 K, under pressure),[9] CeRhSi$_3$ (space group $I4mm$, $T_c \approx$ 1 K, under pressure),[10] CeIrSi$_3$ (space group $I4mm$, $T_c \approx$ 1.6 K under pressure),[11] CeCoGe$_3$ (space group $I4mm$, $T_c \approx$ 0.6 K, under pressure),[12] and Li$_2$Pt$_3$B (space group $P4_332$, $T_c$= 2.2 K).[13–15]

*E-mail address: takeya.hiroyuki@nims.go.jp

Various measuring techniques can be used to probe such unconventional properties. For the particular case of line nodes, power-law relations at lower temperatures are manifested[7,8] in, e.g., NMR relaxation time as $1/T_1 \propto T^3$, penetration depth as $\Delta\lambda \propto T$, thermal conductivity as $\kappa \propto T^3$, and specific heat, which is of relevance to this study, as $C_S(T) \propto T^2$. In particular, $C_S(T)$ of Li$_2$Pt$_3$B does exhibit such a quadratic temperature dependence.[13–15]

It is remarkable that there are only few nonCS strongly correlated electron systems wherein those unconventional superconducting properties were observed (see above). Considering the richness and the diversity of these nonCS-related unconventional properties, then a search for, as well as the characterization of, new nonCS superconductors would be a welcome contribution. Here, we report on the discovery of a new superconducting phase in the ternary Li-Rh-B system. This phase has a nonCS cubic structure and exhibits superconductivity in a wide range of compositions around LiRhB$_{1.5}$. In contrast to the majority of the above-mentioned nonCS superconductors, these Li$_x$RhB$_y$ compounds are nonmagnetic and exibit superconductivity at 2-3 K under ambient pressure.

So as to explore the compositional range of these new superconductors, we synthesized polycrystalline samples of various Li$_x$RhB$_y$ compositions (having the combinations $x$= 0.6, 0.8, 1.0, 1.2, 1.4; $y$=1, 1.25, 1.5, 1.75, 2) by standard solid state reaction of pure elements. After an initial mixing of Rh and B, the starting materials together with a Li lump were placed in a Ta foil or a BN crucible and sealed in a stainless container under an argon atmosphere. Li was handled in an argon-filled glove box. The samples in the stainless container were heated at 700-900 °C for 20 h followed by a furnace cooling. Afterward samples were annealed using the same heat treatment (after being powdered and pelletized).

The prepared samples were black or gray and would show a metallic luster if polished. The weight loss during the heating process was evaluated to be less than 0.2 %. Table I shows the compositions of typical compounds as determined by the Inductively Coupled Plasma (ICP) method. Since lithium is volatile at high temperatures, it is usually difficult to control its@stoichiometry in lithium-based compounds. In addition, it is difficult to determine the precise quantities of the light lithium and boron by using energy-dispersive X-ray analysis or by X-ray diffraction methods. Similarly, it is difficult to ascertain, via the latter method, the site occupancy of Li and B; rather crystal structure determination should be elucidated by neutron diffraction. We believe, nonetheless, that the analyzed data obtained by the ICP method are close to the nominal compositions, that our synthesis process is quite suitable for lithium-rhodium borides, and that the crystal structure determination (specifically the identification of the nonCS space group, see below) is correct.

The crystal structures of all compositions were analyzed by electron as well as X-ray diffraction. Figure 1 shows the electron diffraction pattern along [001] and [110] directions of $Li_{1.2}RhB_{1.5}$. From such results [in particular the extinction rule that the lines $(00l)$ are missing for $l$=odd], the crystal structure is inferred to have a cubic symmetry with space group $P2_13$ or $P4_232$. It is remarkable that both possible space groups have no inversion symmetry (i.e. a nonCS structure).

The X-ray diffraction pattern of the same compound (Fig. 2) confirms the proposed cubic structure having the very same possible space groups (the same extinction rule). Similar electron and X-ray diffraction studies on all studied $Li_xRhB_y$ compounds revealed the same single-phase, cubic crystal structure. In particular, none of the reported Li-Rh-B phases, namely hexagonal $Li_2RhB_2$, orthorhombic $Li_2Rh_3B_2$, $Li_2Rh_3B$ or any rhodium-based borides were detected in the measured diffraction patterns.

Figure 3 shows the $M(T, 20\ Oe)$ curves of $Li_xRhB_y$. Clearly, the superconductivity is evident in most of the studied samples and that their $T_c$ are correlated with the Li/B content. The variation of $T_c$ with the Li/B content (not shown) may be envisaged as being due to an induced variation in the chemical pressure. The influence of the latter can be understood in terms of the BCS relation $T_c = 0.85\theta_D \exp\left(-1/UN(E_F)\right)$ which suggests that pressure could modify $T_c$ via the Debye temperature $\theta_D$, the pairing interaction $U$, or the density of states at the Fermi level $N(E_F)$. Evidently such a correlation of $T_c$ with the volume suggests a similar correlation between $T_c$ and an externally applied pressure $(p)$:

$$\frac{\partial T_c}{\partial p} = \frac{\partial T_c}{\partial V}.\frac{\partial V}{\partial p} = \beta.V.\frac{\partial T_c}{\partial V},$$

where $\beta$ is the compressibility of the solid. To check on this suggestion, we carried out pressure-dependent magnetization of $LiRhB_{1.5}$. Figure 4 shows an extremely weak but positive pressure-dependence of $T_c$: for pressure up to 1 GPa and to a first order in $p$, $T_c \approx 2.5 + 0.15p$ K, $p$ is given in GPa. Though the pressure-dependence of $\beta$, $V$, and lattice anisotropy are unknown, however a simple-minded analysis suggests that such a result can be understood as follows: this particular sample lies within a Li/B region wherein $T_c$ hardly changes with $V$. It is expected that other optimally selected $Li_xRhB_y$ sample would show a more pronounced pressure-dependent effect.

Figure 5 confirms the bulk character of the superconductivity in these $Li_xRhB_y$ samples: on the one hand, the upper panel shows that the maximum shielding fraction of these superconductors amounts to approximately 80% of that of Sn. On the other hand, the lower panel shows the thermal evolution of the electronic specific heat below $T_c$ which is typical of a bulk superconductor (even though there is no discontinuous jump at $T_c$).

The overall behavior of the measured $M$-$H$ isotherms is indicative of a type-II superconducting character and furthermore, based on these curves, the lower critical field $H_{c1}$ was found to follow the relation $H_{c1}=H_{c1}(0)[1-(T/T_c)^2]$ where for Li$_{1.2}$RhB$_{1.5}$, $H_{c1}(0) = 65.6$ Oe and $T_c = 2.5$ K.

The high-temperature resistivities $\rho(T)$ of Li$_x$RhB$_y$ (not shown) exhibit a metallic-like character within the range $T_c(\approx 2.6$ K$) < T \leq 300$. In particular, $\rho(T =300$ K$) \simeq 0.48$ m$\Omega$-cm and the residual resistivity (above $T_c$) is around 0.12 m$\Omega$-cm: thus a residual resistivity ratio of 4. On the other hand, the thermal evolution of the upper critical field, $H_{c2}$ (determined from the midpoint of the superconducting transition of $\rho(T, H)$ curves) is shown in Fig. 6. It happened that due to the limited temperature range of $\rho(T, H)$ measurements, the obtained $H_{c2}$ curve of Li$_{1.2}$RhB$_{1.5}$ can be equally well fitted by the following two functions (consistent with the expected evolution for $T \to T_c$) : either a linear function or the Ginzburg-Landau (GL) formula $H_{c2}(t) = H_{c2}[(1-t^2)/(1+t^2)]$, where $t =T/T_c$. We obtained $T_c = 2.6$ K while $H_{c2}(0)$ equals, respectively, 15.9 kOe and 14.2 kOe; both $H_{c2}$ values are roughly equal but slightly smaller than 19.3 kOe of Li$_2$Pt$_3$B (with almost the same $T_c$).[17] Applying the Ginzburg-Landau (GL) theory with the formula $H_{c2}(t) =\Phi_0/2\pi\xi(0)^2(1$-$t)$ near $T_c$, where $\Phi_0$ is the flux quantum, the GL coherence length $\xi(0) \approx 14.4$ nm, which is roughly similar to that of Li$_2$Pt$_3$B (11.8 nm).[17] The penetration depth $\lambda(0) \approx 14.5$ nm according to the relation $H_{c1}(0) =[\Phi_0/4\pi\lambda(0)^2]\ln[\lambda(0)/\xi(0)]$. Consequently, the GL parameter, $\kappa(0) = [\lambda(0)/\xi(0)]$, is approximately 1.

As evident from above, there is some similarity between the magnetic properties of Li$_x$RhB$_y$ and Li$_2$Pt$_3$B superconductors: both are (Li-B)-based, noncentrosymmetric superconductors and each contains a higher Z atom (Rh and Pt, respectively). As such, the thermal behavior of the superconducting properties of Li$_x$RhB$_y$ might reflect the exotic power-law thermal dependence of Li$_2$Pt$_3$B (see above). For the purpose of checking on the presence of these power-law relations, we studied the thermal evolution of the electronic specific heat contribution of Li$_x$RhB$_{1.5}$ ($x$=0.8, 1.0, 1.2). The measured curves are given in Fig. 5. It is evident that these $C_s(T)$ curves do not follow an exponential evolution. Rather a quadratic-in-temperature law is more appropriate. Accordingly, the above mentioned arguments on the thermal evolution of $C_s$ of a nonCS superconductor suggest that this manifestation of a quadratic-in-temperature behavior of $C_s(T)$ is related to a presence of line nodes in the gap function of Li$_x$RhB$_{1.5}$ (more details on $C_s(T)$ will be reported in Ref.[18]). Finally, it is noted that though some superconducting parameters such as $T_c$ and $\left(\frac{\Delta C}{C_{el}}\right)_{T_c}$ are found to be sample-dependent, the thermal evolution of all $C_s(T < T_c)$ curves is almost the same: this similarity in $\alpha$ is attributed to the observation that the Rh content (which controls the strength of ASOC) is not strongly varied.

In summary, we discovered bulk superconductivity in a wider compositional range of Li$_x$RhB$_y$ with $T_c$ between 2 and 3 K. The structure is inferred to be cubic with a larger $a$-parameter (about 1.21 nm) and a nonCS space group being either $P2_13$ or $P4_232$. The surge of superconductivity within such a nonCS structure elects these compounds as being members of a new family of Li-Rh-B-based nonCS superconductors. The observed correlation between composition, unit-cell volume, and $T_c$ as well as between pressure and $T_c$ suggests a possibility for a further optimization of their superconducting properties: investigations are underway to (i) probe such an optimization route, (ii) to carry out substitution studies on the Li-M-$X$ series (M= transition metal, $X$ =B, As, Si, Ge), and to carry out additional characterization such as penetration depth and NMR so as to provide a further support for the unconventional character of their superconductivity.

**Acknowledgments**

The authors are grateful to the partial financial support of the "Foundation for Promotion of Material Science and Technology of Japan (MST Foundation)". We also acknowledge the partial support received from the Japan Society for the Promotion of Science.

Figure captions

Fig. 1 Electron diffraction patterns of $Li_{1.2}RhB_{1.5}$ using a JEOL JEM-4000EX. The indices on the left bottom of each panel indicate the direction of incidence. Left panel: along [001], right panel: along [110].

Fig. 2 (*Color online*)X-ray diffraction pattern of $Li_{1.2}RhB_{1.5}$. Measurement was collected on a Cu $K_\alpha$ Rigaku RINT-TTR III diffractometer with a Si detector. Symbols represent the measured intensities, solid lines are profile matching patterns calculated by using the so-called profile matching procedure,[16] the short bars represent Bragg positions. The upper (lower) panel compares the measured intensities with the profile matching pattern using the space group $P4_232$ ($P2_13$). These space groups were determined based on the extinction rule [absence of (00$l$) lines for $l$=odd].

Fig. 3 (*Color online*) $M(T, 20\ Oe)$ of upper:(a) $Li_xRhB_{1.0}$, middle:(b) $Li_xRhB_{1.5}$, and lower:(c) $Li_xRhB_{2.0}$, all measured within the range of $1.8 \leq T \leq 3.2$ K under a zero-field-cooling (ZFC) condition. Bulk samples were cut into a cylindrical shape ($\phi$0.45 x0.50 cm), sealed in a gelatin capsule (all handled in an inert-gas glove box), and subsequently measured on a superconducting quantum interference device (SQUID) magnetometer (Quantum Design MPMS-5S). The double superconducting transitions observed in the magnetization curves of some samples are attributed to the granular character of these particular samples.

Fig. 4 (*Color online*) Pressure-dependent magnetization curves of $LiRhB_{1.5}$. Inset: the evolution of $T_c$ with the applied pressure can be approximated as $T_c \approx 2.5 + 0.15p$ K ($p$ in GPa). Pressure effect was measured with a low-temperature hydrostatic micropressure cell (up to around 1 GPa) operated within a SQUID magnetometer environment. Daphne oil was used as a pressure-transmitting fluid while Sn as a manometer.

Fig. 5 (*Color online*) Upper: ZFC magnetizations ($H$=20 Oe) of $Li_{1.2}RhB_{1.0}$, $Li_{0.8}RhB_{1.5}$, and $Li_{1.0}RhB_{2.0}$. The shielding volume fractions are compared with a similar-sized Sn sample (cylinders of $\phi$0.45x0.50 cm). Lower: thermal evolution of the electronic specific heat of $Li_xRhB_{1.5}$ ($x$=0.8, 1.0, 1.2). Measurements were carried out on a semi-adiabatic calorimeter. For each sample, the usual Debye coefficient $\beta$ and Sommerfeld coefficient $\gamma$ were obtained from the analysis of the standard plot $C/T$ versus $T^2$ within the normal-state region (not shown). Afterwards, the value of the calculated phonon contribution ($\beta T^3$) is subtracted from the total measured specific heat to give the electronic contribution $C_{el}(T)$ (shown in this graph as symbols). The thin lines represent the calculated normal-state electronic

contribution ($\gamma T$). As the shielding fraction is found to be ~80% (see the upper panel), then the calculated $C_s$ (shown as thick line) is considered to be a sum of ~20% of the normal-state contribution expressed as $x\gamma T$ and ~80% of the quasi-particle contribution expressed as $(1-x)\alpha (T/T_c)^2$ where $x = 0.2$ and $\alpha$ is a temperature independent coefficient (which measures the strength of the ASOC). The inset table shows the values of $\alpha$ (in mJ/moleK), $\beta$ (in mJ/moleK$^4$), and $\gamma$ (in mJ/moleK$^2$).

Fig. 6 (*Color online*) Left panel: magnetoresistance curves of Li$_{1.2}$RhB$_{1.5}$ sample (0.11x0.12x0.45 cm$^3$) measured under 1 mA on a conventional four-points method using a home-made probe in a MPMS-5S environment. Right panel: $H_{c2}(T)$ curves of Li$_{1.2}$RhB$_{1.5}$ as determined from the midpoint of the superconducting transition of the magnetoresistivity curves. $H_{c2}(0)$ was analyzed by a linear fit (solid line) and also by the GL fit (dashed line) using the formula $H_{c2}(t) = H_{c2}[(1-t^2)/(1+t^2)]$, where $t = T/T_c$.

Table I. Representative compositions as determined by the Inductively Coupled Plasma (ICP) method. Before the start of the determination, we used aqua regia first and then $K_2S_2O_7$ to completely dissolve Rh.

| Material | Analyzed composition (atomic ratio) | |
| --- | --- | --- |
| (nominal composition) | Li/Rh | B/Rh |
| $Li_{0.8}RhB_{1.0}$ | 0.81 | 1.01 |
| $Li_{0.8}RhB_{1.5}$ | 0.87 | 1.47 |
| $Li_{1.0}RhB_{1.5}$ | 0.95 | 1.48 |
| $Li_{1.0}RhB_{2.0}$ | 0.99 | 1.96 |

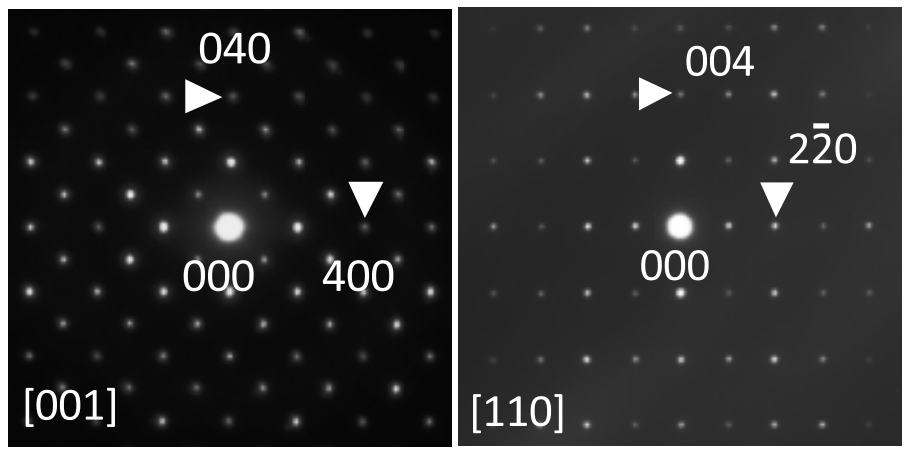

Fig. 1. Takeya *et al (2010)* Electron diffration

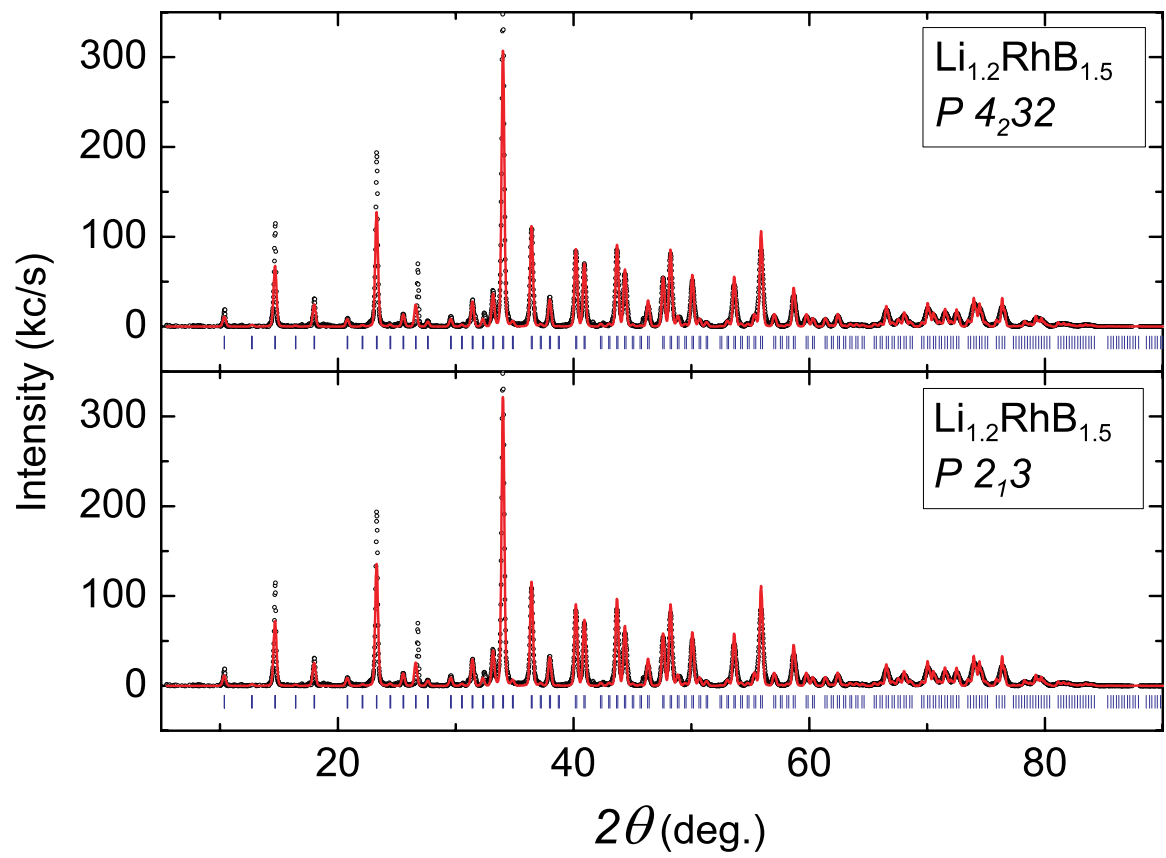

Fig. 2. Takeya *et al (2010)* - Xrd Diff

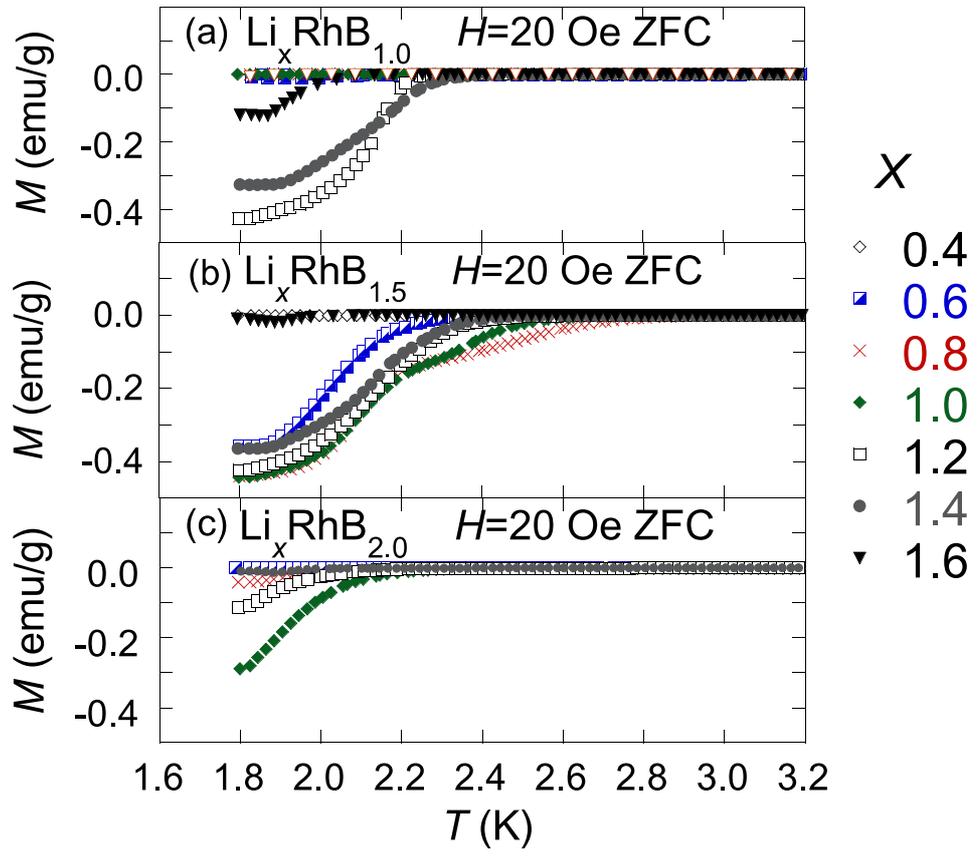

Fig. 3. Takeya *et al (2010)* MvsT

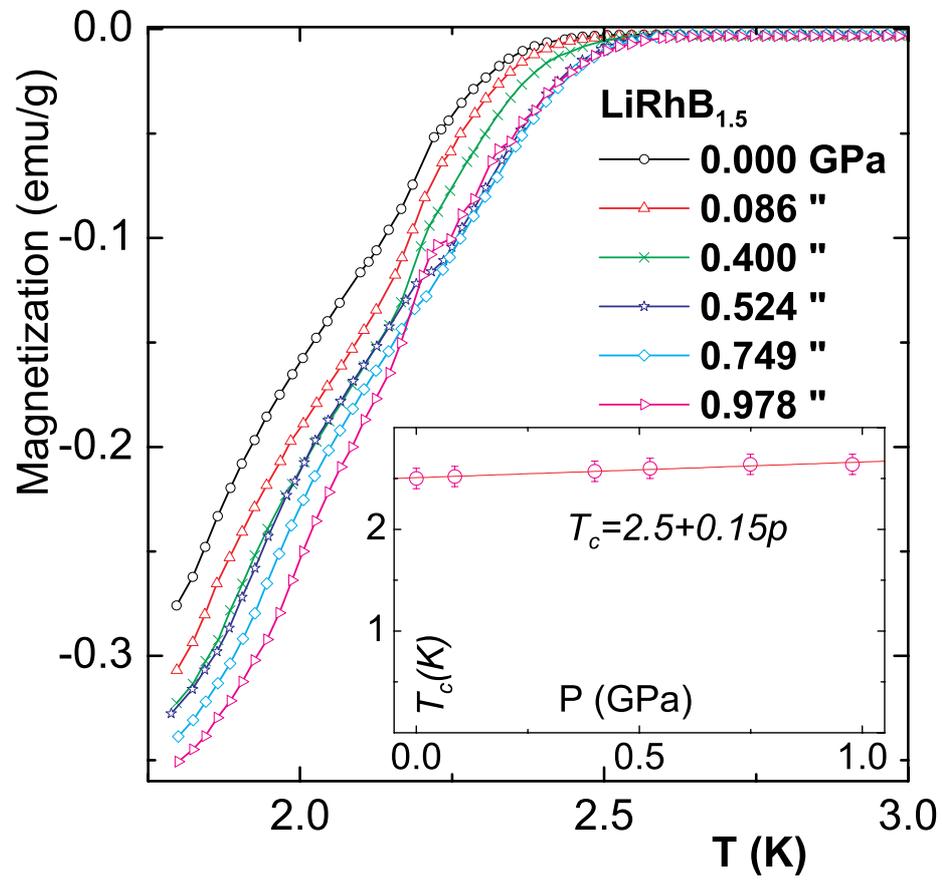

Fig. 4. Takeya *et al (2010)* Magnetization under Pressure

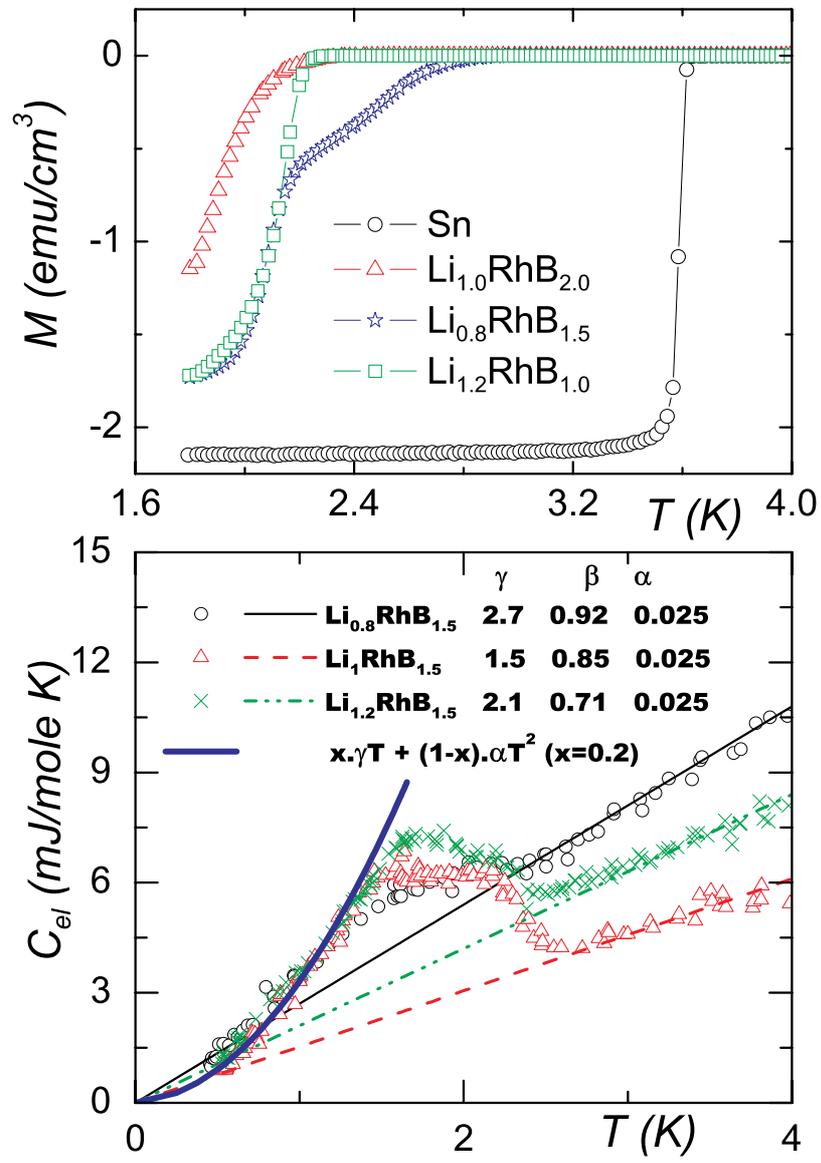

Fig. 5. Takeya *et al (2010)* Shield Fraction and specific heat

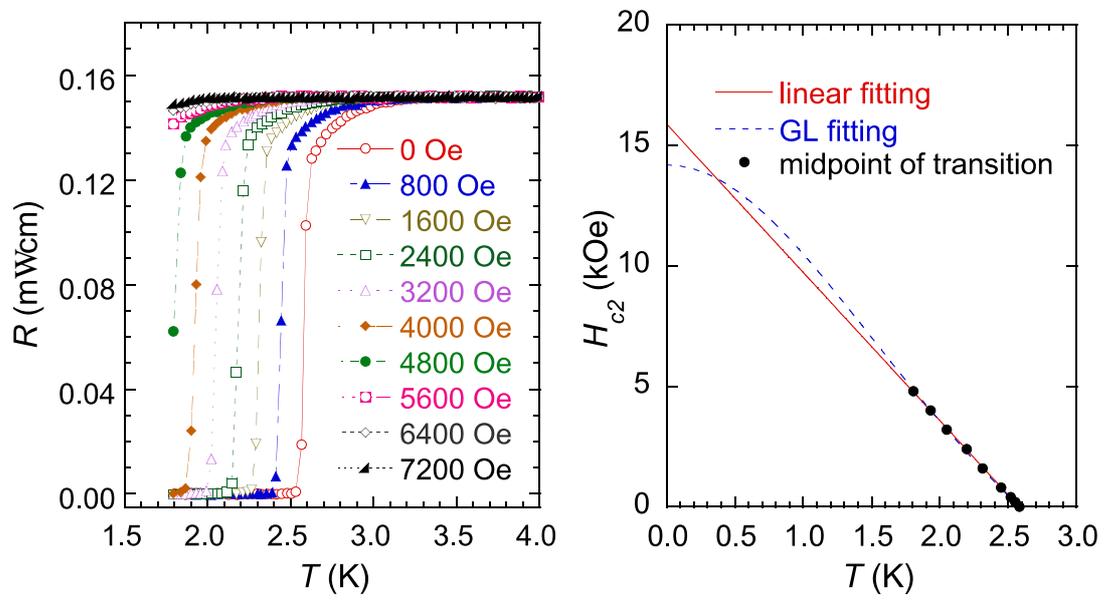

Fig. 6. Takeya *et al (2010)* RvsT, HcvsT